\documentclass[aps,nofootinbib,preprint]{revtex4}
\usepackage{amsfonts,amssymb,amsmath}
\usepackage{graphicx}
\usepackage{booktabs}
\def\d{\delta}
\def\g{\sqrt{-g}}

\newcommand{\f}[2]{\frac{#1}{#2}}
\newcommand{\mk}[1]{\left( #1 \right)}
\newcommand{\kk}[1]{\left[ #1 \right]}

\newcommand{\be}{\begin{equation}}
\newcommand{\ee}{\end{equation}}
\newcommand{\DE}{{\rm DE}}
\newcommand{\lcdm}{{\rm \Lambda CDM}}

\begin{document}

\title{
Future Oscillations around Phantom Divide in $f(R)$ Gravity
}

\author{
Hayato Motohashi$^{~1,2}$, Alexei A. Starobinsky$^{~2,3}$,
and Jun'ichi Yokoyama$^{~2,4}$
}

\address{
$^{1}$ Department of Physics, Graduate School of Science,
The University of Tokyo, Tokyo 113-0033, Japan \\
$^{2}$ Research Center for the Early Universe (RESCEU),
Graduate School of Science, The University of Tokyo, Tokyo 113-0033, Japan \\
$^{3}$ L. D. Landau Institute for Theoretical Physics,
Moscow 119334, Russia \\
$^{4}$ Institute for the Physics and Mathematics of the Universe(IPMU),
The University of Tokyo, Kashiwa, Chiba, 277-8568, Japan
}

\begin{abstract}
It is known that scalar-tensor theory of gravity admits regular
crossing of the phantom divide line $w_{\DE}=-1$ for dark energy,
and existing viable models of present dark energy for its
particular case -- $f(R)$ gravity -- possess one such crossing in
the recent past, after the end of the matter dominated stage. It
was recently noted that during the future evolution of these
models the dark energy equation of state $w_{\DE}$ may oscillate
with an arbitrary number of phantom divide crossings. In this
paper we prove that the number of crossings can be infinite,
present an analytical condition for the existence of this effect
and investigate it numerically. With the increase of the present
mass of the scalaron (a scalar particle appearing in $f(R)$
gravity) beyond the boundary of the appearance of such
oscillations, their amplitude is shown to decrease very fast. As a
result, the effect quickly becomes small and its beginning is
shifted to the remote future. 
\end{abstract}

\begin{flushright}
RESCEU-1/11
\end{flushright}

\maketitle

\section{Introduction}

The accelerating expansion of the present Universe is confirmed by current precise
observational data such as type Ia supernovae~\cite{Perlmutter:1998np,Riess:1998cb},
anisotropy of cosmic microwave background~\cite{Komatsu:2010fb}, large scale
structure~\cite{Tegmark:2003ud} and baryon acoustic oscillations~
\cite{Eisenstein:2005su,Percival:2007yw}. The standard cosmological constant
($\rm \Lambda$)-Cold-Dark-Matter (${\rm CDM}$) model is indeed able to explain these
observational results within observational errors. In this model a cosmological
constant is regarded as a new fundamental physical constant. However, the required value
of the cosmological constant is very tiny compared with any known physical scales. Thus, its
relation to the standard quantum theory of known particles and fields is not understood
today although some nonperturbative effects may naturally generate such a small quantity,
see e.g. Refs.~\cite{Yokoyama:2001ez,Kiefer:2010pb}. More generally, a source of the
current cosmic acceleration is called dark energy (DE). Further we shall use the more
detailed term "present DE" for it to distinguish it from
primordial DE which was responsible for another accelerated expansion regime, dubbed
inflation, which occurred in the very early Universe~
\cite{Starobinsky:1980te,Sato:1980yn,Guth:1980zm}. The relation of primordial DE to
the known elementary particles has not been established, too.

Both primordial and present DE can have either a physical origin (some new physical fields
of matter) or a geometrical one. In the latter case, the Einstein gravity becomes modified.
One of the simplest and self-consistent generalizations of the Einstein gravity is $f(R)$
gravity which incorporates a new phenomenological function $f(R)$ of the Ricci scalar $R$
(with $d^2f/dR^2$ not identically zero) into the action, see Eq.~\eqref{action} below. For a
long time this theory of gravity
was known to contain viable inflationary models, among them the simplest one introduced in
Ref.~\cite{Starobinsky:1980te} that remains in agreement with the most recent observational data.
Thus, $f(R)$ gravity can successively describe primordial DE. Rather recently, after many
unsuccessful trials, viable models of present dark energy were found
\cite{Hu:2007nk,Appleby:2007vb,Starobinsky:2007hu} which provide non-trivial alternatives to
the standard $\lcdm$ model.\footnote{In order not to destroy the standard early Universe
cosmology, including the recombination, the correct Big Bang nucleosynthesis and inflation
of any kind, these models of present DE possessing a non-trivial form of $f(R)$ in the low-$R$,
$R>0$ region have to be further generalized by changing the behaviour of $f(R)$ at large $R$
and by extending it to the region of negative $R$, see Ref.~\cite{Appleby:2009uf}. However, this
generalization is not important for our study.}
This theory is a special class of the scalar-tensor theory of gravity with the vanishing
Brans-Dicke parameter $\omega_{BD}$. It contains a new scalar degree of freedom dubbed
"scalaron" in Ref.~\cite{Starobinsky:1980te}, thus, it is a {\em nonperturbative} generalization
of the Einstein gravity. From the quantum point of view, scalaron is a massive spin-$0$ particle
which mass depends on $R$. We consider $f(R)$ gravity as a phenomenological macroscopic theory of
gravity, alternative to the Einstein one, without discussing its microscopic origin.\footnote{Note
the simplest possible mechanism that has attracted a new interest recently: a scalar field $\phi$
with some potential and the non-minimal coupling $-\xi R\phi^2/2$ to the Einstein gravity in the
limit of a very large negative $\xi$ (i.e. the sign of coupling is opposite to that of the
conformally coupled case). However, this mechanism leads to $df/dR > 1$. So, while sufficient to
produce the functional form of $f(R)$ needed for successful inflationary models, it is not useful
for construction of viable models of present DE.}

The existence of this additional degree of freedom imposes a number of constraints
on the functional form of $f(R)$ in viable cosmological models. In particular, in order to
have the correct Newtonian limit, as well as the standard matter-dominated stage
with the scale factor behaviour $a(t)\propto t^{2/3}$ driven by cold dark matter and baryons,
the following conditions should be fulfilled for $R\gg R_0$ where $R_0\equiv R(t_0)\sim H_0^2$,
$t_0$ is the present moment and $H_0$ is the Hubble constant, and up to curvatures in the
centre of neutron stars:
\be |f(R)-R|\ll R,~~|f'(R)-1|\ll 1,~~Rf''(R)\ll 1. \ee
Here the prime denotes the derivative with respect to the argument $R$. Furthermore, $f(R)$
should satisfy the following conditions to guarantee both that Newtonian gravity solutions
are stable and that the standard matter-dominated Friedmann-Robertson-Walker (FRW) stage remains
an attractor with respect to an open set of neighbouring generic cosmological solutions in $f(R)$
gravity:
\be f'(R)>0,~ f''(R)>0. \ee
In quantum language, the first condition means that gravity is attractive and graviton is not
a ghost, while the second one -- that scalaron is not a tachyon.
Specific functional forms of $f(R)$ satisfying these conditions, as well as laboratory and Solar
system tests of gravity, and possessing a future stable (or at least metastable) de Sitter stage
that is required for correct description of observable properties of present DE, have been
proposed in Refs.~\cite{Hu:2007nk,Appleby:2007vb,Starobinsky:2007hu},
and much work has been carried out on their cosmological consequences.

In order to describe the difference between FRW background solutions of $f(R)$ gravity and the
$\lcdm$ model, it is useful to introduce the effective equation-of-state~(EoS) parameter for DE
$w_{\DE}\equiv P_{\DE}/\rho_{\DE}$ where the effective pressure $P_{\DE}$ and the effective energy
density $\rho_{\DE}$ of DE are determined using the Einsteinian representation of gravitational
field equations, see Eqs.~\eqref{E1},~\eqref{E2} below. Another independent parameter which
describes scalar (density) perturbations on a FRW background is the gravitational growth index
$\gamma$ defined as $d\ln \d/d\ln a\equiv \Omega_m(z)^{\gamma(z)}$ where $\d\equiv \d\rho_m/\rho_m$
and $\Omega_m\equiv 8\pi G\rho_m/3H^2$ are a matter density fluctuation and the density parameter
for matter, respectively. In $f(R)$ gravity, $w_{\DE}$ is time dependent and $\gamma$ is time and
scale dependent whilst they keep the constant value $w_{\DE}=-1$ and $\gamma\approx 6/11$ in the
$\lcdm$ model. Time and scale dependency of $\gamma$ generate an additional transfer function for
matter density fluctuation that constrains the model parameter region~\cite{Motohashi:2010tb,Motohashi:2009qn,Motohashi:2010sj}.

One of the most interesting features of geometrical DE distinguishing it from physical DE based
on non-ghost physical fields minimally coupled to gravity, like quintessence, is the possibility
of phantom behaviour, $w_{\DE}<-1$, of DE. Moreover, this behaviour may well be temporary with DE
becoming normal, $w_{\DE}> -1$, after smooth crossing of the phantom boundary $w_{\DE}=-1$. In
particular, models of geometrical DE based on scalar-tensor gravity were long known to admit this
property~\cite{BEPS00}. $f(R)$ gravity is a particular case of scalar-tensor gravity, so it permits
phantom behaviour of DE and smooth crossing of the phantom boundary, too. Existing observational
data do not exclude the possibility of phantom behaviour of DE (though they do not specifically
favour it, too) for the following simple reason: as was noted above, DE in the particular form of an
exact cosmological constant ${\rm \Lambda}$ is in a good agreement with all data. But since
$w_{\rm \Lambda}\equiv -1$, it lies exactly at the phantom boundary. Thus, any small deviation of DE
from ${\rm \Lambda}$ to the direction of decreasing $w_{\DE}$ results in its phantom behaviour. So,
theory has to be prepared for this possibility that explains large interest in DE models admitting
it. However, it is clear already that this "phantomness" should be small. In particular,
if it is assumed for simplicity that $w_{\DE}={\rm const}$, when $|w_{\DE}+1|<0.1$ at the
approximately $2\sigma$ confidence level~\cite{Komatsu:2010fb}.

Moreover, viable $f(R)$ models of present DE~\cite{Hu:2007nk,Appleby:2007vb,Starobinsky:2007hu}
generically exhibit phantom behaviour during the matter-dominated stage and one recent crossing of
the phantom divide $w_{\DE}=-1$ even in the case of the smoothest behaviour of a FRW scale factor
$a(t)$, when there were no superimposed small oscillations of $a(t)$ in the past (in quantum language,
no condensate of primordial scalarons with the zero momentum)
~\cite{Hu:2007nk,MMA09,Appleby:2009uf,Motohashi:2010tb}. From the physical point of view, the absence
of primordial scalarons in the viable $f(R)$ models of present DE is needed in order not to destroy
the standard early Universe cosmology and it can be achieved by primordial inflation of any kind,
see Ref.~\cite{Appleby:2009uf} for a detailed discussion. Using numerical calculations, it has been
recently shown that even in this smoothest case the EoS parameter $w_{\DE}$ can oscillate around the
future de Sitter solution in these DE models~\cite{Bamba:2010iy}, see also Ref.~\cite{LKM10}.

To investigate the phenomenon of multiple crossing of the phantom divide in more detail and
analytically, in the present paper we prove that this crossing can indeed occur infinitely many times
during the future evolution of viable $f(R)$ models of present DE if the scalaron mass at a future
stable de Sitter stage in these models is sufficiently large. Though this phenomenon is not directly
observable since it refers to remote future, it is interesting and important from the theoretical
point of view. Also, it is possible to check from observational data at the present moment if the
derived analytical criterion for the existence of the infinite number of crossings is satisfied or
not.

Thus, the present paper focuses on the oscillatory behaviour of $w_{\DE}$ around the phantom divide
$w_{\DE}=-1$ in the future. In Sec.~II, we review the stability conditions and the condition of the
existence of a stable future de Sitter stage in $f(R)$ gravity, and derive the condition for the
existence of the infinite number of these oscillations analytically using the perturbation theory
around the de Sitter solution. In Sec.~III, we focus on the specific viable model of present DE in
$f(R)$ gravity and present results from numerical calculations relating this condition to observable
properties of the Universe at the present time. Sec.~IV is devoted to conclusions and discussion.

\section{The criteria}

The action studied is of the form
\be S= \frac{1}{16\pi G} \int d^4x \g f(R) + S_m, \label{action}\ee
where $f(R)$ is a function of the Ricci scalar $R$ and $S_m$ denotes matter action with the minimal
coupling to gravity. Field equations are derived as
\begin{align}
&R_{\mu\nu}-\f{1}{2}g_{\mu\nu}R=8\pi G(T_{\mu\nu} + T^\DE_{\mu\nu})~, \label{eq1} \\
&8\pi GT^\DE_{\mu\nu}=(1-F)R_{\mu\nu}-\f{1}{2}(R-f)g_{\mu\nu}+( \nabla_\mu \nabla_\nu -
g_{\mu\nu} \square ) F
\label{eq2}
\end{align}
where $F=df/dR$. We use the representation \eqref{eq1},\eqref{eq2} to define the effective
energy-momentum tensor $T^\DE_{\mu\nu}$ of DE. The $(0,0)$ and $(i,i)$ components of the field
equations are
\begin{align}
3FH^2 &= \f{RF-f}{2} - 3H\dot F + 8\pi G \rho, \label{energyeq}\\
6F\f{\ddot a}{a} &= RF-f -3(\ddot F +H\dot F) -8\pi G (\rho+3P).
\end{align}
It is also useful to use the trace equation:
\be \label{tre} RF-2f+3\square F=8\pi GT. \ee
Thus, the effective energy density, the pressure and the EoS parameter of DE have the form:
\begin{align}
8\pi G \rho_{\DE} &\equiv 3H^2-8\pi G\rho= -3(1-F)\f{\ddot a}{a} + \f{R-f}{2} - 3H\dot F, \label{E1}\\
8\pi G P_{\DE} &\equiv -2\dot H-3H^2-8\pi GP= (1-F)\mk{\f{\ddot a}{a}+2H^2}-\f{R-f}{2} +\ddot F +
2H\dot F,
\label{E2} \\
w_{\DE}+1 &= \f{2(1-F)(-\ddot a/a+H^2)+\ddot F -H\dot F}{-3(1-F)\ddot a/a + (R-f)/2 - 3H\dot F}.
\end{align}

During an asymptotic de Sitter regime, the matter density decreases rapidly as
$\rho\propto e^{-3H_1t}$ and soon can be neglected. Therefore, it follows from Eq.~\eqref{tre} that
a constant value of the Ricci scalar $R=R_1=\rm{const}=12H_1^2$ at a de Sitter regime is given by a
root of the algebraic equation
\be \label{dsc} 2f_1=R_1F_1 \ee
where $f_1\equiv f(R_1)$ and $F_1\equiv F(R_1)$. At the de Sitter regime, DE is characterized by
$8\pi G \rho_{\DE,1} = -8\pi G P_{\DE,1} = \f{R_1}{4}$, thus $w_{\DE,1} = -1$.

To investigate the stability of the future de Sitter solution and the possibility of oscillatory
behaviour around it, we expand Eqs.~\eqref{energyeq},~\eqref{tre} in the perturbation series with
respect to $\d R \equiv R-R_1$ and $\d H\equiv H-H_1$. In the first order in $\d R$ and $\d H$,
\begin{align}
&\d H=-\f{H_1F_{R1}}{2F_1}(\d R'-\d R)+\f{1}{2F_1H_1}\f{8\pi G \rho_m}{3}~, \label{dH} \\
&\d R''+3\d R'+\f{1}{3H_1^2}\mk{\f{F_1}{F_{R1}}-R_1}\d R=\f{8\pi G \rho_m}{3F_{R1}H_1^2}~, \label{dR}
\end{align}
where prime denotes the derivative with respect to the number of $e$-folds $N\equiv \ln a=-\ln (1+z)$
and $F_{R1}\equiv F_R(R_1)\equiv dF(R_1)/dR$. Although the matter density term in the right-hand side
has the zero order, we include it because $\rho_m\propto e^{-3H_1t}$ is much smaller than background
quantities at the future de Sitter stage.

Eq.~\eqref{dR} is solved as a sum of the homogeneous solution of Eq. \eqref{dR} with the zero
right-hand side, $\d R_{\rm osc}$, and the special solution of the non-homogeneous equation
$\d R_{\rm dec}$:
\be \d R = \d R_{\rm dec}+\d R_{\rm osc}. \ee
Since $\rho_m=\rho_{m0}e^{-3N}$, $\d R_{\rm dec}$ is obtained as
\be \d R_{\rm dec}=\f{8\pi G\rho_{m0}}{F_1-R_1F_{R1}}e^{-3N}. \ee
Notice that it describes a monotonically decaying mode.

On the other hand, the homogeneous solution $\d R_{\rm osc}$ may have decaying, growing and
oscillatory behaviour. In the case of monotonic behaviour (both roots of the characteristic equation
for Eq.~\eqref{dR} are real), to keep the future de Sitter solution stable (a stable node), the
coefficient of the third term in the left-hand side of Eq.~\eqref{dR} should be positive. So, the
stability condition is  \cite{MSS88}:
\be \label{stc} \f{F_1}{F_{R1}}>R_1~. \ee
The oscillatory behaviour occurs when the de Sitter asymptote is a focus (complex roots). For this,
the discriminant of the characteristic equation should be negative:
\be \label{onc} \f{F_1}{F_{R1}}>\f{25}{16}R_1~. \ee
Since the coefficient of the second term in the left-hand side of Eq.~\eqref{dR} is positive, the
focus is always stable and the inequality \eqref{onc} is stronger than \eqref{stc}. The criterion
\eqref{onc} of the oscillatory approach to the future de Sitter asymptote is equivalent to the
condition \be M_1^2 \equiv \f{F_1- R_1F_{R1}}{3F_{R1}} > \f{9H_1^2}{4} = \f{3R_1}{16}~, \ee
where $H_1, R_1$ and $M_1$ are the Hubble parameter, the scalar curvature and the scalaron mass at
the future de Sitter state. If the oscillation condition is satisfied,
\be \d R_{\rm osc}=Ae^{-3N/2}\sin (\omega N+\phi) \ee
where $\omega\equiv 2\sqrt{\f{F_1}{R_1F_{R1}}-\f{25}{16}}$, and $A$ and $\phi$ are integration
constants.

The perturbation of the EoS parameter $\d w_{\DE}=(\d P_{\DE}+\d \rho_{\DE})/\rho_{\DE,1}$
is calculated from $8\pi G(\rho_\DE+P_\DE)=-2\dot H-8\pi G\rho_m$ and Eq.~\eqref{dH},
\be \d w_{\DE}=\f{4}{R_1}\kk{-\f{R_1F_{R1}}{3F_1}\d R'+\f{1}{3}\mk{\f{R_1F_{R1}}{F_1}-1}\d R+
\mk{\f{4}{F_1}-3}\f{8\pi G \rho_m}{3}}. \ee
We decompose $\d w_{\DE}\equiv \d w_{\rm dec}+\d w_{\rm osc}$ as
\begin{align}
\d w_{\rm dec}&=\f{4}{R_1}\mk{\f{1}{F_1-R_1F_{R1}}-1}8\pi G \rho_{m0}(1+z)^3\\
\d w_{\rm osc}&=A(1+z)^{3/2}\f{4}{R_1}\kk{-\f{R_1F_{R1}}{3F_1}\omega\cos(\omega N+\phi)+\f{1}{3}
\mk{\f{5R_1F_{R1}}{2F_1}-1}\sin(\omega N+\phi)}.
\end{align}
Thus, in the latter case of a stable de Sitter solution with oscillations, $w_{\DE}$ crosses the
phantom boundary $w_{\DE}=-1$ infinitely many times during the future evolution of the Universe.

\section{The specific model}

We consider the following viable $f(R)$ model~\cite{Starobinsky:2007hu}:
\be f(R)=R + \lambda R_s \kk{\mk{1+\f{R^2}{R_s^2}}^{-n}-1}, \ee
where $n$ and $\lambda$ are model parameters, and $R_s$ is determined by the present observational
data, namely, the ratio $R_s/H_0^2$ is well fit by a simple power-law $R_s/H_0^2=c_n\lambda^{-p_n}$
with $(n,c_n,p_n)=(2,4.16,0.953),~(3,4.12,0.837),$ and $(4,4.74,0.702)$,
respectively~\cite{Motohashi:2010tb}.

From Eq.~\eqref{dsc}, the equation for de Sitter solutions is
\be \alpha(r)\equiv r+2\lambda\kk{\f{1+(n+1)r^2}{(1+r^2)^{n+1}}-1}=0, \ee
where $r\equiv R_1/R_s$.
It is obvious that the Minkowski space-time, $r=0$, is one of the solutions.
We denote the other positive solutions for $\alpha(r)=0$ as $r_a\equiv R_{1a}/R_s$ and
$r_b\equiv R_{1b}/R_s$. $r_a$ and $r_b$ can be estimated by considering the limits of small and
large $r$. For $r\ll 1$, $\alpha(r)\simeq r[1-\lambda n(n+1)r^3]$, and for $r\gg 1$,
$\alpha(r)\simeq r-2\lambda$. Therefore, the de Sitter solutions are
$r=r_a\simeq [\lambda n(n+1)]^{-1/3}$ and $r=r_b\simeq 2\lambda$. Strictly speaking, these
approximations are valid either for $\lambda \gg 1$ or, in the case of $r_a$, for $n\gg 1$ (while
$\lambda$ may be of the order of unity). However, it follows from the numerical analysis that the
solutions for $n=2$ and $\lambda=3$ are already close enough to these analytical estimations.

One can check their stability and oscillatory behaviour using the stability parameter $\beta(r)$ and
the oscillation parameter $\gamma(r)$ which are derived from Eqs.~\eqref{stc} and \eqref{onc}:
\begin{align}
\beta(r)&\equiv \f{(1+r^2)[(1+r^2)^{n+1}-2n\lambda r]}{2n\lambda[(2n+1)r^2-1]}-r >0, \label{beta}\\
\gamma(r)&\equiv \f{(1+r^2)[(1+r^2)^{n+1}-2n\lambda r]}{2n\lambda[(2n+1)r^2-1]}-\f{25}{16}r >0.
\label{gamma}
\end{align}
Since $\gamma(r)=\beta(r)-9r/16$, there is no oscillations for an unstable de Sitter solution, as it
should be. From these criteria we see that $r=r_a$ and $r=r_b$ are an unstable de Sitter solution
and a stable de Sitter solution, respectively. The specific values are presented in the
Table~\ref{tbrb}.

\begin{figure}[h]
\centering
\includegraphics[width=110mm]{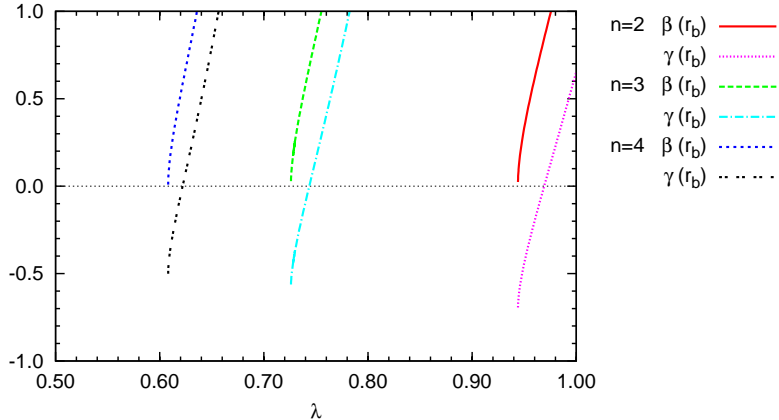}
\caption{Values of the stability parameter $\beta$ and the oscillation parameter $\gamma$ for stable
de Sitter solutions for various model parameters. The parameter regions $\gamma(r_b)<0<\beta(r_b)$ and
$\gamma(r_b)>0$ correspond to stable de Sitter solutions without oscillations and with oscillations,
respectively.}
\label{bg}
\end{figure}

For a fixed $n$ and various values of $\lambda$, we obtain $\lambda_\beta$ and $\lambda_\gamma$ as
roots of $\beta(r_b)=0$ and $\gamma(r_b)=0$, respectively. Models are classified according to
$\lambda$ being in the intervals $\lambda<\lambda_\beta,~\lambda_\beta<\lambda<\lambda_\gamma$, and
$\lambda>\lambda_\gamma$, and in each region a de Sitter solution $r=r_b$ is unstable, stable
without oscillations, and stable with oscillations, respectively. Although most of the parameters
realize a stable de Sitter solution with oscillations (a stable focus), there exists a parameter
region corresponding to a stable de Sitter solution without oscillation (a stable node).
Fig.~\ref{bg} suggests that such parameter regions are $0.944<\lambda<0.970$, $0.726<\lambda<0.744$
and $0.608<\lambda<0.622$ for $n=2,~3$ and $4$, respectively.

\begin{table}[t]
\centering
\caption{Stable de Sitter solutions for various model parameters $n$ and $\lambda$.
$r_b\equiv R_{1b}/R_s$ is a stable de Sitter solution in terms of the normalized Ricci scalar.
$\beta$ and $\gamma$ denote the stability~\eqref{beta} and oscillation~\eqref{gamma}
parameters, respectively.}
\label{tbrb}
\begin{tabular}{rr|lll} \toprule
$n$ & $\lambda$ & $r_b$ & $\beta(r_b)$ & $\gamma(r_b)$ \\ \hline
2 & 1 & $1.64$ & $1.58$ &    $6.56\times 10^{-1}$ \\
2 & 3 & $5.99$ & $8.54\times 10^2$ &    $8.51\times 10^2$ \\
2 & 10 &    $20.0$ & $3.23\times 10^5$ &    $3.23\times 10^5$ \\
3 & 1 & $1.94$ & $1.37\times 10$ &  $1.26\times 10$ \\
3 & 3 & $6.00$ & $1.53\times 10^4$ &    $1.53\times 10^4$ \\
3 & 10 &    $20.0$ & $6.17\times 10^7$ &    $6.17\times 10^7$ \\
4 & 1 & $1.99$ & $5.07\times 10$ &  $4.95\times 10$ \\
4 & 3 & $6.00$ & $3.31\times 10^5$ &    $3.31\times 10^5$ \\
4 & 10 &    $20.0$ & $1.44\times 10^{10}$ & $1.44\times 10^{10}$ \\
\bottomrule
\end{tabular}
\end{table}

\begin{figure}[t]
\centering
\includegraphics[width=75mm]{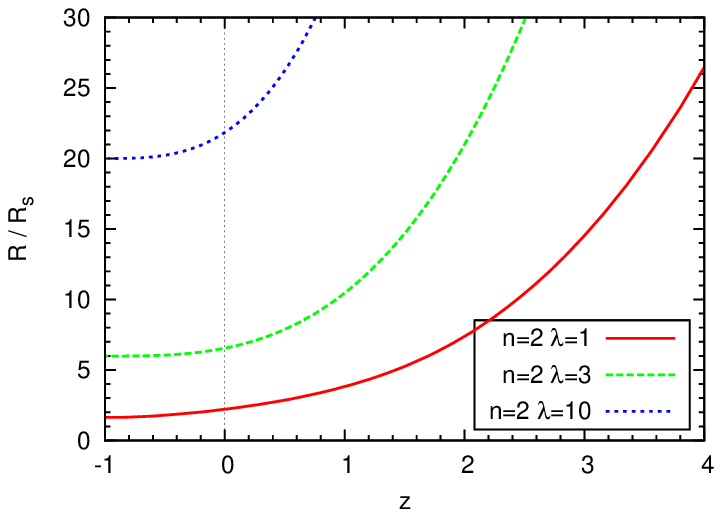}
\includegraphics[width=75mm]{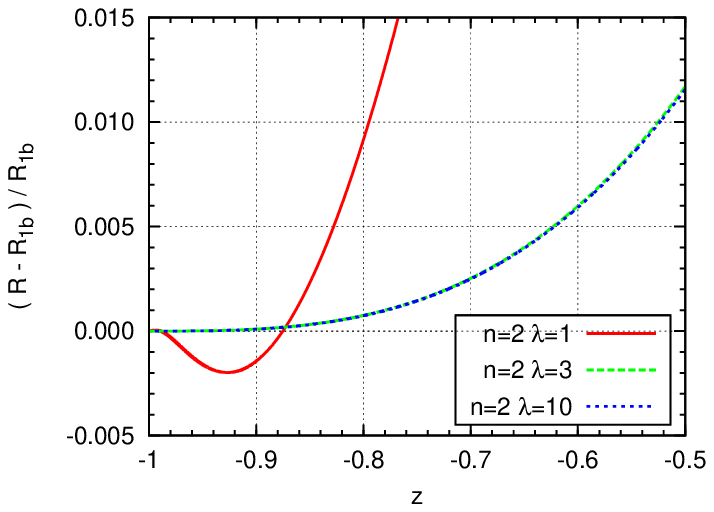}
\caption{Future evolution of the Ricci scalar. It approaches the stable de Sitter solution which is
presented in the Table~\ref{tbrb}.}
\label{R}
\end{figure}

\begin{figure}[t]
\centering
\includegraphics[width=75mm]{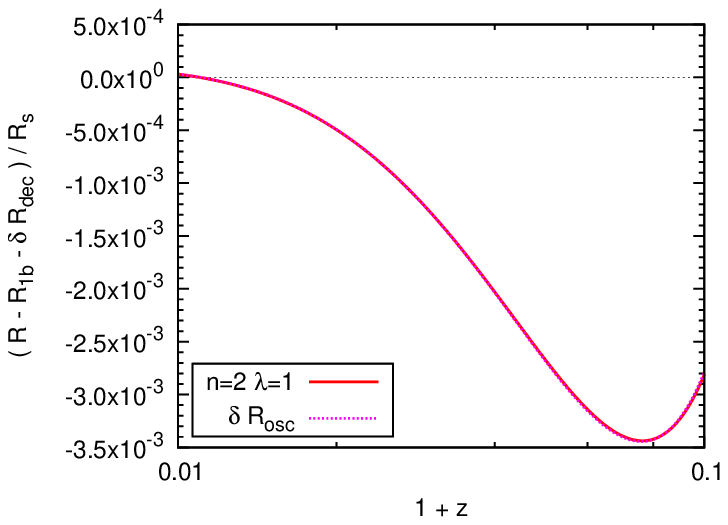}
\includegraphics[width=75mm]{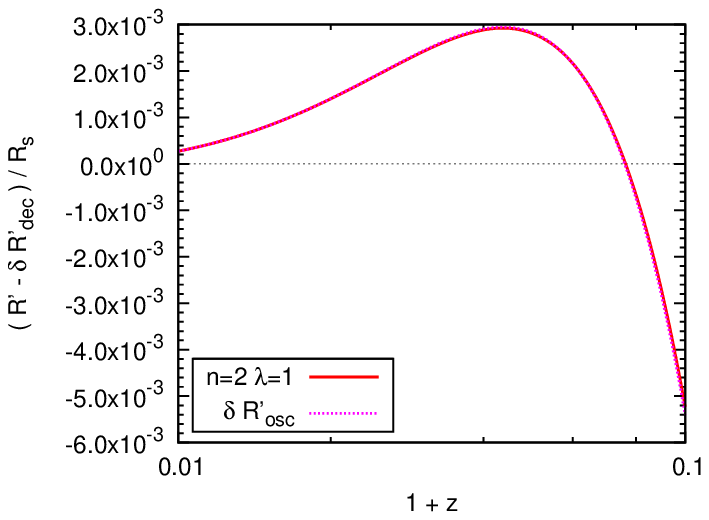}
\includegraphics[width=75mm]{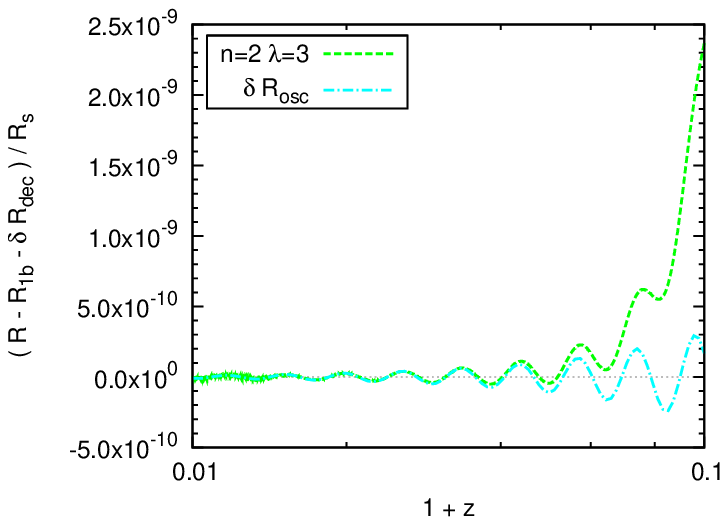}
\includegraphics[width=75mm]{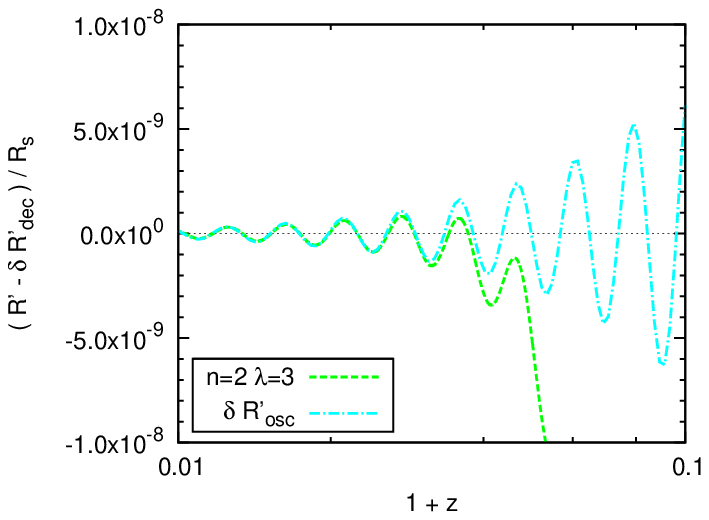}
\caption{Numerical results for $\d R-\d R_{\rm dec}$ compared with the analytic solution for
$\d R_{\rm osc}$.}
\label{rsub}
\end{figure}

We integrate the evolution equations numerically.
Initial condition are set at $z=10$ using the $\lcdm$ model, and the present moment is determined by
the condition $\Omega_m=0.27$. Fig.~\ref{R} shows that $R$ approaches a stable de Sitter solution.
It is seen from the right panel of Fig.~\ref{R} that the perturbation theory with respect to
$\d R\equiv R-R_{1b}$ is valid when $z\lesssim-0.8$ for $n=2,~\lambda=1$, and when $z\lesssim-0.5$
for $n=2,~\lambda=3$ or $10$. The oscillation of $\d R$ for $n=2,~\lambda=1$ is clearly seen. For
$n=2$ and $\lambda=3$ or $10$, oscillations exist, too, but their amplitude is so small that we cannot
see them. To make them visible, we have subtracted the decaying mode $\d R_{\rm dec}$ in
Fig.~\ref{rsub}. The analytic solution for $\d R_{\rm osc}$ fits the result well.

\begin{figure}[t]
\centering
\includegraphics[width=75mm]{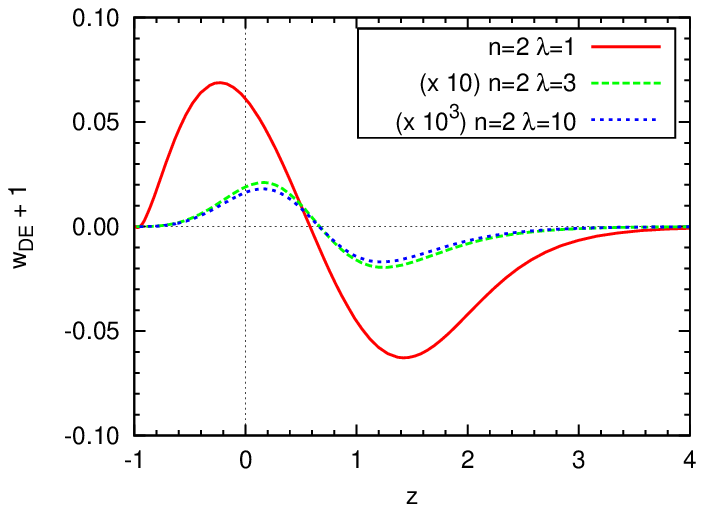}
\includegraphics[width=75mm]{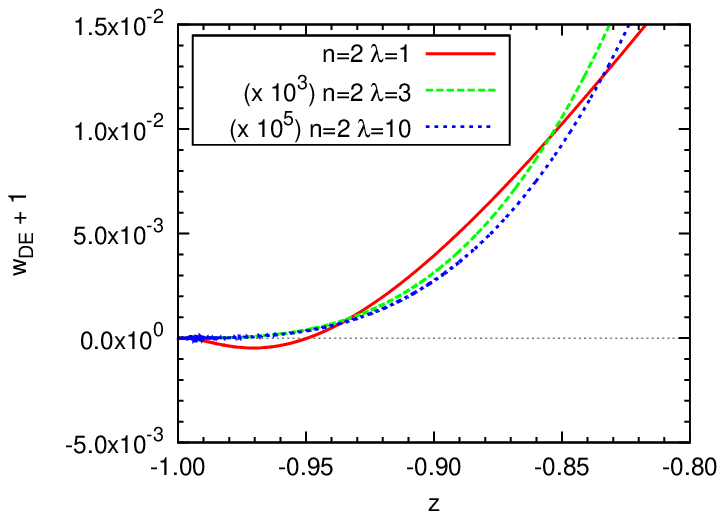}
\caption{Future evolution of the effective EoS parameter for dark energy.}
\label{w1310}
\end{figure}

\begin{figure}[t]
\centering
\includegraphics[width=75mm]{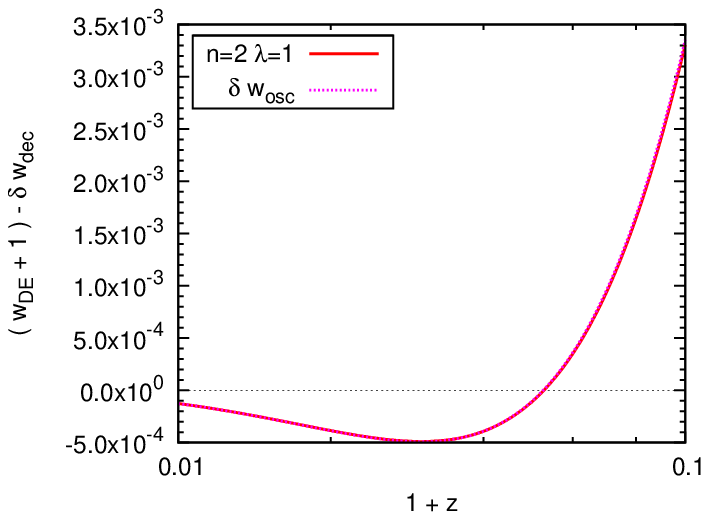}
\includegraphics[width=75mm]{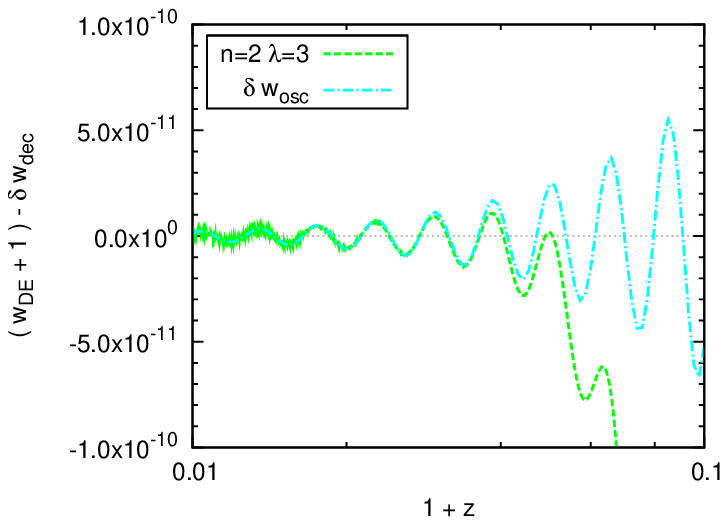}
\caption{Numerical results for $(1+w_{\DE})-\d w_{\rm dec}$ compared with the analytic solution for
$\d w_{\rm osc}$.}
\label{wzw}
\end{figure}

\begin{figure}[t]
\centering
\includegraphics[width=75mm]{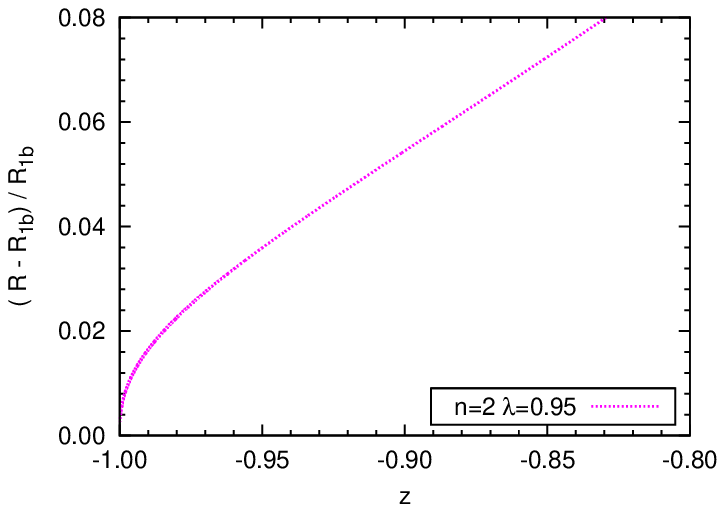}
\includegraphics[width=75mm]{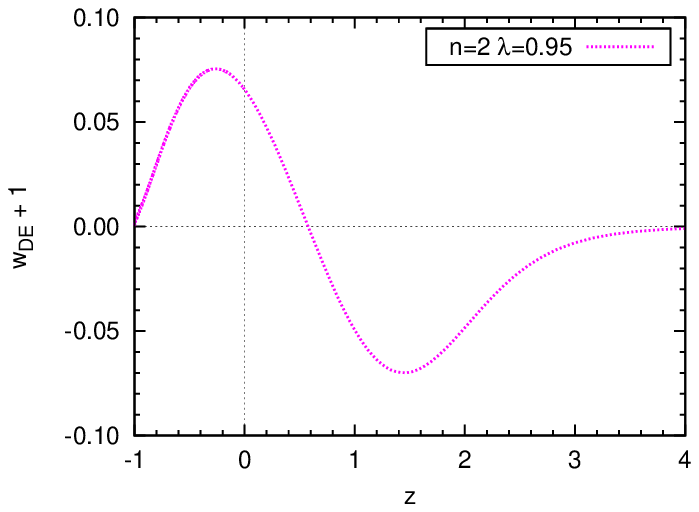}
\includegraphics[width=75mm]{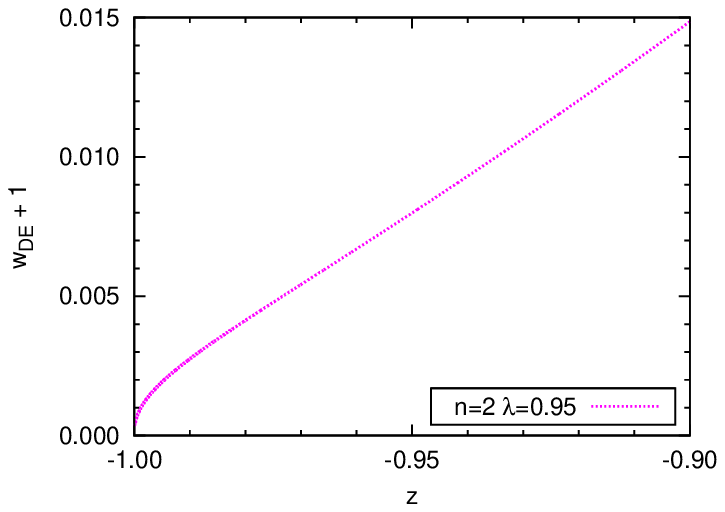}
\caption{Future evolution of the effective EoS parameter for dark energy for $n=2$ and $\lambda=0.95$.
There is no oscillations.}
\label{rw0.95}
\end{figure}

Fig.~\ref{w1310} depicts the evolution of the EoS parameter for $n=2$ and $\lambda=1,~3,~10$.
The first phantom crossing occurred in the past at $z\sim 0.5$ in agreement with
Ref.~\cite{Motohashi:2010tb}. We subtract the decaying mode $\d w_{\rm dec}$ and present the
oscillation mode $\d w_{\rm osc}$ in Fig.~\ref{wzw}. The numerical results are very close to the
analytic solutions for $n=2,~\lambda=1$ and $3$. For $n=2,~\lambda=10$, the amplitude of the
oscillations is small and the frequency is large, so that we cannot distinguish them from numerical
noise. Finally, we present the case $n=2,~\lambda=0.95$ in Fig.~\ref{rw0.95} as an example of a
non-oscillatory approach to the stable de Sitter solution. Note that the trajectories of $\d R$ and
$\d w$ are convex upward and there is no oscillations indeed.

\section{Conclusions and discussion}

We have investigated conditions under which the effective EoS parameter $w_\DE$ of present DE in
$f(R)$ gravity can oscillate an infinite number of times around the phantom boundary $w_{\DE}=-1$
during the future evolution of the Universe.
The analytical condition of the existence of this phenomenon, Eq.~\eqref{onc}, is derived that
depends on the properties of $f(R)$ near a future stable de Sitter stage only. The physical sense of
this condition is that the rest mass of the scalaron (a massive scalar particle which arises in $f(R)$
gravity in addition to massless spin-2 graviton) should be sufficiently large at the future de Sitter
stage. Thus, this phenomenon is generic. However, the amplitude of these oscillations has been shown
to decrease fast with the increase of the scalaron mass beyond the boundary of the appearance of such
oscillations. As a result, the effect quickly becomes small and its beginning is shifted to the
remote future. For real scalaron masses lying below this boundary, the future stable de Sitter stage
is reached without the phantom boundary crossing. Analytic solutions for the behaviour of $w_{\DE}$
near the phantom boundary have been obtained in the first order of the small quantity $|w_{\DE}+1|$.
Generically they have a monotonically decaying part $\d w_{\rm dec}$ and a damped harmonic
oscillatory part $\d w_{\rm osc}$. For a specific viable $f(R)$ model of present DE energy, numerical
integration of FRW background evolution has been performed which future behaviour is in a good
agreement with the analytic formulas.

All calculations have been done for the smoothest initial conditions in the past corresponding to the
absence of a primordial homogeneous oscillating scalaron component. So, even in this case, an oscillating
scalaron component (the condensate of scalarons with the zero momentum) arises around the present moment
when the scalaron mass is comparable to the Hubble constant $H_0$ (in the units where $\hbar = c=1$),
and it quickly becomes dominant over the non-relativistic matter component (cold dark matter and
baryons) in the future. But its effective energy-momentum tensor in turn soon becomes negligible
compared to an effective cosmological constant producing the future stable de Sitter stage. For less
smooth initial conditions, more phantom boundary crossings may occur in the past. But these initial
conditions are hardly compatible with the standard cosmology of the early Universe confirmed by
numerous observational data. We hope to return to this question elsewhere.

Finally, though the very phenomenon of multiple (and even infinite) number of phantom boundary crossings
in the future is not directly observable, it is very interesting and important from the theoretical
point of view. Also, as follows from our numerical calculations of the full evolution from
the remote past to the remote future, the scalaron mass at the future de Sitter stage is close to its
present value. Therefore, in principle it is possible to check from observational data describing
the present and the past of our Universe if the derived analytical criterion for the existence of an
infinite number of oscillations in $w_{\DE}$ is satisfied or not.

\acknowledgments{
AS acknowledges RESCEU hospitality as a visiting professor. He was
also partially supported by the grant RFBR 08-02-00923 and by the
Scientific Programme ``Astronomy'' of the Russian Academy of
Sciences. This work was supported in part by JSPS Research
Fellowships for Young Scientists (HM), JSPS Grant-in-Aid for
Scientific Research No.\ 19340054 (JY), Grant-in-Aid for
Scientific Research on Innovative Areas No. 21111006 (JY), JSPS
Core-to-Core program ``International Research Network on Dark
Energy'', and Global CE Program ``the Physical Sciences
Frontier'', MEXT, Japan.
}

\end{document}